# AN ICA-ENSEMBLE LEARNING APPROACH FOR PREDICTION OF UWB NLOS SIGNALS DATA CLASSIFICATION

Jiya A. Enoch[1], Ilesanmi B. Oluwafemi[2], Francis A. Ibikunle[1] and Olulope K. Paul[2]

**Correspondence**: jiya.adama@lmu.edu.ng [1]Department of Electrical and Information Engineering, Landmark University, Omu-Aran, Nigeria

**Abstract:** Trapped human detection in search and rescue (SAR) scenarios poses a significant challenge in pervasive computing. This study addresses this issue by leveraging machine learning techniques, given their high accuracy. However, accurate identification of trapped individuals is hindered by the curse of dimensionality and noisy data. Particularly in non-line-of-sight (NLOS) situations during catastrophic events, the curse of dimensionality may lead to blind spots due to noise and uncorrelated values in detections. This research focuses on harmonizing information through wireless communication and identifying individuals in NLOS scenarios using ultra-wideband (UWB) radar signals. Employing independent component analysis (ICA) for feature extraction, the study evaluates classification performance using ensemble algorithms on both static and dynamic datasets. The experimental results demonstrate categorization accuracies of 88.37% for static data and 87.20% for dynamic data, highlighting the effectiveness of the proposed approach. Finally, this work can help scientists and engineers make instant decisions during SAR operations.

## I. Introduction

Around the globe, many natural disasters with variable frequencies occur, including the destruction of man-made infrastructure like bridges and buildings, earthquakes, fire mishaps, wildfires, floods, tsunamis, and volcanic activity. The remaining people in such a situation are liable to become entrapped in the openings left by the collapsing building materials. The presence, whereabouts, and proportion of trapped people are unknown to search and recovery workers during post-disaster rescue efforts (Anyfantis & Blionas, 2021).

These catastrophic calamities have killed millions of individuals in the past few decades. The effectiveness of post-disaster rescue actions and assistance to victims has already become essential in reducing the anticipated fatalities, as the deadliest natural catastrophes have tended to precede one another. The creation of effective techniques that enable trustworthy robotic navigation and monitoring of uncharted settings is still an ongoing study area in the realm of security and protection (Aparna et al., 2020 ).

The identification of human targets via walls is crucial for counterterrorism, surveillance, safety, and emergency recovery efforts. Due to its many benefits, including high definition, great penetrability, high positioning accuracy, low power consumption, and a lack of sensitivity to channel degeneration, ultra-wideband (UWB) radar is perfect for piercing barriers to detecting hidden objects. Because the radar can pass through materials like buildings, it can identify targets hiding beneath them as well as track, locate, and identify other hidden targets. For law enforcement authorities, it has a variety of applications. Target identification, range-based localization, health tracking, and person detection via the wall are just a few of the areas where UWB has now excelled (Wang et al., 2021).

Because of the rapid advancement of machine learning, particularly deep learning, professionals can now instantly make the best decisions based on goal-oriented data using a high-performance computer program that extracts high-level characteristics from multi-dimensional large data. Methods for facilitating disaster relief and humanitarian relief are among the most popular areas of data science (Terranova et al., 2021).

The authors (Zhan et al., 2016) propose a downstream channel assessment network that relies on CNNs for feature extraction and continuous networks for channel prediction. Nonetheless, such techniques didn't specifically center on NLOS detection. The majority of the aforementioned methods train a classifier to recognize the NLOS using features taken from the CIR. Although such algorithms work effectively in reality, their feature extraction somehow doesn't take into account all of the data in the CIR.

In this research, the ensemble classifier and recognition in the domain of through-the-wall person detection are used using an ICA approach in machine learning feature extraction. We introduce Standardscaler, a data normalizer algorithm with a rigorous training procedure. After analyzing the dataset's assessment methods and the data pre-processing, trials were run to categorize and detect the human targets while behind walls in different stages.

The goal of this research is to employ an ICA for (static and dynamic) feature extraction and an ensemble algorithm for categorization and identifying human victims hidden behind barriers. Also, a comparative analysis between the two datasets (static and dynamic) was carried out to evaluate their performance. The rest of this research is structured as follows: The second section concentrates on the associated work's guiding concepts. The creation and training of the data are discussed in the third section, along with the materials and technological approaches. The testing approach and analysis of the results are described in the fourth section. The fifth portion contains the conclusion.

## II. Disaster management in search and rescue of humans in NLOS detection

Through disaster management, it is possible to prepare for, react appropriately to, recuperate from, and mitigate damages from disasters. Emergency management includes preventing catastrophes before they happen, responding to disasters right away, and assisting with and reconstructing societies after the calamity. Emergency management is essential to everybody's security and ought to be taken into account in all everyday decisions rather than simply in the event of a tragedy, which is becoming all too frequent (Pour, 2021).

Robust communication networks are essential for effective emergency management activities following a disaster, whether it is natural or man-made. Large-scale catastrophic situations, unfortunately, may destroy telecommunications networks and hinder search and rescue efforts (Ilbeigi et al., 2022).

Due to the expansion of humanitarian activities and requirements, it is essential to resolve the difficulties presently seen on the ground. Delays, traffic, poor connectivity, and unaccountability are problems that could provide testing grounds for the purported benefits of emerging technological innovations (Rodríguez-Espíndola et al., 2020).

The entire management system is seriously threatened by the existing dependence on centralized, physical infrastructure approaches. Additionally, present procedures for keeping in touch when systems are inoperable mainly rely on employing interim infrastructures, such as telecommunications towers. The primary focus of the whole procedure is handling assistance demands premised on current disaster information and effectively addressing those requests by allocating the management's accessible restricted resources (Samir et al., 2019).

Blockchain-based technologies facilitating search and rescue (SAR) operations could be built with the most intelligent functionalities as information and communications technology (ICT) and drones, including the Internet of Things (IoT), cloud technologies, image analysis, and unmanned drones, improve (Nguyen et al., 2021).

Nevertheless, IoT, blockchain, and crowdsourcing techniques can provide useful information that will make the provisioning procedure go more smoothly. To support this improvement, they may also serve as a way to create a dynamic mutual trust among those who offer assistance, those who seek it out, and those who give it. Additionally, it can serve as a foundation for integrating technologies like blockchain, 3D printing, and artificial intelligence to enhance the transfer of data, goods, and financial means in humanitarian distribution chains (Rodríguez-Espíndola et al., 2020).

## III. Literature review

Utilizing machine learning algorithms to recognize and find people entrapped beneath fallen structures has received considerable interest and study over the last two decades. A blockage along the line of sight (LOS) between a sender and receiver, on the other hand, has a negative influence on the results of straightforward radio signal measurements, such as the Received Signal Strength Indicator (RSSI) values (Rosli et al., 2019), making them unsuitable for the scenarios being investigated.

Lately, Yu et al. (2017) have suggested a method that uses higher-order cyclo stationarity to identify a human's breathing and pulse. The vital signs were found using the third-order cyclic cumulant. The harmonic intermodulation, erratic body movements, and clutter noise could be reduced with this technique. Additionally, it enabled the radar sensor to pick up on weak signals with low SNR levels. Target surveillance cantered on the radar sensor is also crucial in emergency response or military activities, in addition to the vital sign's detection methods.

To follow various objects behind a wall, a method centered on an algorithm named variational mode decomposition (VMD) was used. The respiratory impulses were divided up into different sub-signals using this technique. The VMD algorithm can detect split signals. Calculate traveled respiratory recognition, distance bins, VMD technique, and Hilbert transform were the four processes that made up the tracking algorithm. The outcome of the suggested algorithm was contrasted with the traditional FFT. The VMD algorithm technique is a good option for different targeted tracking using a microwave radar device (Yan et al., 2016)

The second group focuses on machine learning methods to improve generalization capacity. When dealing with various NLOS conditions brought on by various settings, like wall materials and object orientation. Patterns that might be challenging to find using traditional methods can be successfully analyzed using ML methodologies. One of the first methods to categorize and evaluate human respiratory activity with machine-learning WiFi sensors was also developed as part of the work (Khan et al., 2017). In this research, Khan et al. developed a convolutional neural network (CNN), which had a 94.85% precision rate.

Suárez-Casal and José (2019), have published a thorough overview of the most current research on UWB through-wall radar that takes into account signal-processing techniques like individual vital sign tracking and theories of human detection. The researchers recognized three important approaches for human detection (i.e., establishing a person's existence or absence in the examined radar data) using NLOS sensors: (i) the constant false alarm rate (CFAR), whose objective is to establish the energy level within which any response can be assumed to the most likely come from a target as compared to any of the erroneous channels; (ii) the numerical properties of the received signal (also used in the work are skewness, kurtosis, energy, etc.); and (iii) the multipath designs of the radar reflected signal.

For unspecified scenarios, Park et al. (Park et al., 2020) suggested an Ultra-wideband NLOS detection based on the transfer of learning strategy with precision comparable to deep learning algorithms trained with 30 data points and 48 execution periods. Nevertheless, they did not examine other designs or human detection tasks; they only examined Convolutional neural systems and multidimensional perceptions.

The research in (Lazaro et al., 2014) utilized two receiving antennas to find the vital signs in a manner comparable to this. Nevertheless, they only decided to analyze the powerful signal. The chosen signal was then subjected to the mobile filter to eliminate the quasi-static noise. In (Chioukh et al., 2014), harmonic radar was proposed as a further method to lessen flicker interference. This study proposes a dual-frequency harmonic CW radar sensing system (fundamental and harmonic). They asserted that this strategy significantly lowered flicker interference while increasing SNR.

When creating a microwave radar device to find living people under debris, there are several challenges to overcome. First, when the phase shift caused by the victim's range from the detector is an even multiple of 2, the zero-point phenomenon in the microwave radar detector manifests. The signal that was received in this instance is close to zero. Second, distortion from motion artifacts or interference from several people reduces the microwave radar sensor's ability to detect vital signs (Peng et al., 2016). The distortion from an operator near the radar antennae and the clutter in the shadowing region are two other noticeable interferences that the electromagnetic search and rescue device must contend with. This research examines this significant radar and communication technology. The sample radar sensor technologies for exploration and recovery are described in this article and are centered on two broad types of radar sensors: continuous wave (CW) and ultra-wideband (UWB) technologies. Initially, a theoretical foundation and an examination of the literature on radar systems are provided. Then, many types of radar sensors for search and rescue missions are covered, including their functioning concepts and physical construction. The conclusion and outlook for the future are offered at the end.

The rest of this paper is organized as follows: A review of the associated literature is in Section II. The materials and data set creation process are given in Section III. The classification methods and the process for extracting the features are covered in Section IV. In Section IV, the outcomes of the suggested approaches for accuracy improvement are shown. In Section V, the findings are presented.

## VI. Materials and Method

The main idea of this research is to forecast machine learning tasks on high-dimensional UWB NLOS data for human identification in lower-dimensional datasets. The datasets comprised of both static and dynamic data were employed in this research. A summary proposed framework is shown in Figure 1. The procedure is modified by the step of relevant data extraction from a specific dataset using the ICA feature extraction technique for both static and dynamic datasets. The output of the Ensemble classification algorithm is compared to existing methods to assess the performance of the NLOS dataset.

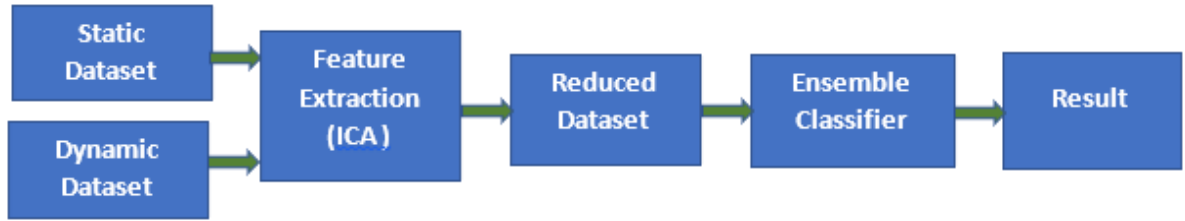

Figure 1. Proposed Framework

To examine and identify NLOS human detection data of various heterogeneous materials, numerous machine learning algorithms have been replicated. The necessity of examining dataset collection and methods employing particular data mining techniques are discussed (Moro et al., 2022). Numerous studies conducted by researchers in this field are consulted, and recent reviews of studies looking at the NLOS dataset are done. By employing ensemble classifier evaluation techniques to rate enormous sets of data acquired with NLOS, a supervised machine learning method for NLOS human detection was proposed. The standard scaler displays reduced data by utilizing dimensionality reduction techniques.

A smaller and more refined set of data is displayed in the output for improved prediction. Results from the training of the person detection dataset showed the effectiveness of a supervised hidden learning-based feature extraction method and underlined the necessity of data collection techniques for human detection analysis. For the classification of the UWB NLOS dataset using a supervised model, a generalized method of extremely accurate data classification was presented, incorporating an unbiased collection of condensed dimensions and space feature extraction techniques. On human-detectable features, barriers, objects with varied orientations, and at various distances from the gathered datasets, a standard scaler was applied.

The work by (Moro et al., 2022) executed a comprehensive estimation effort in various settings and targeted discrepancies to generate a repository for implementing and analyzing machine learning algorithms as human classifiers. They used the Novelda NVA-R661 radar development kit, which operates in the 6-8.5 GHz range and is based on the NVA-6201 chip.

Data from experiments were gathered in various indoor settings on the foremost floor of the Cesena Campus of the University of Bologna's School of Engineering. The static measurement, which involves keeping the radar stationary on a moving cart at a distance of roughly 130 cm from the ground, was first taken into account. The radar (r) was positioned at distances of 30, 60, and 90 cm from the target (d), which was put 20 cm behind the obstacles. The material thickness was also thought of as a potential hindrance. The following materials are specifically taken into account when gathering data: double glazing set at 10 cm, wooden doors set at 3 and 5 cm, brick wall set at 15 cm, and a glass window set at 2 cm. The second instance provides a more realistic circumstance in which the radar was handed at various heights, making the acquisition dynamic through minor movements (Moro et al., 2022).

As a result, the strategy used in this research can be summed up as follows:

1) Create a dataset of UWB NLOS from various victims' body orientations and materials of different barriers.
2) Show how data normalization affects the correctness of the dataset.
3) Select the latent component from the dataset using ICA feature extraction methods.
4) Outline a strategy for classifier training that utilizes ensemble classification to improve prediction accuracy.
5) Using the ensemble classifier to suggest classification, and
6) comparing the results to other methods using relevant literature.

## IV.I. Methods

The data acquired from (Moro et al., 2022) were evaluated experimentally using MATLAB, and features were extracted using ICA. Utilizing an ensemble algorithmic technique, classification was carried out using retrieved features.

The research utilizes standard scalar preprocessing, known as SC, to standardize acquired data, remove noisy values and outliers, perform data transformation, and normalize the data. SC aims to convert various eigenvalues into a predetermined range of 0s and 1s. This approach employs item scaling to ensure features are roughly equal in size, making them relatively important and easier for machine learning algorithms to process. The Standard Scaler, a part of SC, standardizes features by subtracting the average and scaling the variance to one, thus achieving unit variance by dividing all values by the standard deviation. This standardization, also called normalization, is particularly useful for data with a Gaussian distribution, facilitating easier management by machine learning methods. Unlike normalization, standardization doesn't have a bounding range, potentially affecting outliers in the data. However, normalizing data may not always be necessary (Sarra et al., 2022).

The ensemble is used to dynamically optimize the features and the classifiers while learning the representation of the features. The loading of the dataset, feature extraction, classification, and outcome modules make up the proposed system. The feature extraction unit loads the dataset from the human detection module, which has been normalized, and applies the ICA algorithm to it. The output module is delivered to the feature extraction module, which uses feature extraction independently. After the response has been accurately classified by the classification algorithm utilizing Ensemble, it is then delivered in the outcome module. In the context of NLOS data analysis, SC feature selection is paired with the feature extraction algorithms "ICA" and "Ensemble."

These methods will be merged and applied to the creation of an evaluation metrics model. The following is the article's methodology: Using ensemble classification, SC data normalization, and ICA feature extraction to improve classification performance, we compare our results concerning accuracy, specificity, sensitivity, and precision.

## IV.I. Materials

To improve the efficiency of the human detection dataset, this paper investigates independent component analysis (ICA) and the ensemble classification technique for dimensionality reduction of high-dimensional NLOS data.

Supervised learning algorithms, which typically accept labeled instances as input, are the learning procedures utilized following the structure of the training dataset. As a result, the class of concern will be identified for each occurrence. To enable the final model to categorize ensuing pre-processed waveforms into one of the following situations, waveform instances in this case are labeled as either "Person YES" or "Person NO."

This investigation takes into account ICA and ensemble classification techniques to significantly reduce the dimensionality of human detection data. Two datasets are presented as an I-J matrix M. Total instances for the static scenario are I = 23,552; for the dynamic case, I = 17,408. J = 256 (K) is applied in both scenarios. the total number of rows produced by adding together each pulse set from the measurement experiment, removing the incorrect elements (i.e., those with low information content and high distortion levels) following a qualitative examination (Moro et al., 2022). Information obtained from the University of Bologna's Cesena Campus School of Engineering.

Table 1 provides a detailed summary of the dataset, which consists of occurrences of the feature samples as well as sampling properties.

Table 1. Features of the dataset

| Dataset Attributes | Description |
|---|---|
| Radar Dynamic | 23,552 |
| Radar Static | 17,408 |
| Observation | 256 |
| Source | University of Bologna's Cesena Campus School of Engineering |
| Characteristics | Human body orientation, building materials, debris, and radar distances |
| Accessibility | Openly accessible dataset (Moro et al., 2022) |

**IV.III. Dimensionality Reduction**

Dimensionality reduction approaches are vital techniques used in statistical analysis, machine learning, and computational complexity to address the challenges posed by high-dimensional datasets. These techniques aim to reduce the number of primary variables under study, thereby mitigating the curse of dimensionality. Before employing unsupervised methods like clustering algorithms, dimension reduction is often applied as a preprocessing step (Imperial, 2019).

Dimensionality reduction serves several purposes: it reduces the computational time and space required, simplifies the interpretation of input variables for machine learning algorithms by eliminating multicollinearities, and enables data visualization in lower-dimensional spaces such as 2D or 3D

In machine learning, dimensionality reduction involves two main processes: feature extraction and feature selection. Feature extraction involves identifying and extracting related features from high-dimensional data to create a smaller set of informative features. This process enhances the interpretability of the data and reduces redundancy and noise. Feature selection, on the other hand, involves choosing a subset of features that best represent the data while discarding unnecessary or duplicated features. (Chakraborty et al., 2022).

**IV.VI. Feature Extraction**

Large amounts of raw data are divided into more manageable groupings using a dimensionality reduction approach called feature extraction. These enormous data sets are similar in that they contain a lot of components that require a lot of processing power. The term "feature extraction" refers to techniques that select relevant variables and/or integrate them to create features, reducing the amount of data that must be processed while also properly and completely defining the initial data set (Karabacak & Özmen, 2022).

For decreasing sizes of high-dimensional data, feature extraction is a clever replacement for feature selection. It is called "feature translation or creation" in a lower-dimensional domain. By altering the initial parameter in a lower-dimensional space, the feature extraction method represents problems in a more useful and discriminating space, thereby enhancing the efficiency of subsequent analysis. There are two primary types of feature extraction algorithms: linear and non-linear techniques. Compared to non-linear procedures, linear approaches are typically quicker, more reliable, and easier to understand. Non-linear approaches identify intricate data structures, or embedments, that linear methods are unable to discern (Salman, 2010).

By converting the data into a more straightforward form of characteristics, feature extraction is utilized from a particular dataset, to extrapolate more latent optimal component features. It provides an open data depiction of the corresponding parameter that is used to aggregate linear parameters into feature subsets. Moreover, feature extraction is a general-purpose technique that can be done using a variety of methods (Shafizadeh-Moghadam, 2021). This paper uses ICA for system-matching linked parameters since it involves an orthogonal transformation with models of continuously uncorrelated elements.

## IV.V. Independent Component Analysis (ICA)

ICA was initially presented in the 1980s and proposed a reiterative instantaneous technique. The suggested technique was not relevant in a variety of situations, and no theoretical justification was provided in that publication. However, the ICA technique was mostly unknown until 1994, when the term "ICA" first surfaced and was promoted as a novel idea (Tharwat, 2020).

ICA seeks to harvest from the data relevant information or underlying signals (a set of measured mixture signals). ICA was utilized to recover source signals. ICA is sometimes thought of as a dimensionality reduction process when it can retain or eliminate a particular source. This procedure, also called a filtering operation, can remove or filter certain information.

ICA enhances higher-order metrics like kurtosis and can discover independent components. There are numerous ICA algorithms, including Infomax and a FastICA projection pursuit (Liu et al., 2021). Identifying separate components through these strategies primarily aims to minimize mutual information, maximize non-Gaussianity, or use the maximum likelihood (ML) estimate approach (Agrawal et al., 2022).

---

**Algorithm 1** ICA

---

1: Set the value of K to zero.
2: Calculate the distance between the input sample and the training instances.
3: Sort the separation in step three.
4: Select the top K-nearest neighbors in step four.
5: Apply the simple majority in step five.
6: Use more neighbors to determine the loaded sample's class label.

End

---

## IV.IV. Classification

Recent trends in data analysis highlight the innovative use of ranked probabilistic representations, derived from both line-of-sight (LOS) and non-line-of-sight (NLOS) data, to address classification challenges. This approach, as discussed by (Rayavarapu & Mahapatro, 2022), involves refining these representations using a combined framework, enabling the creation of effective classification algorithms. Meanwhile, advancements in machine learning have spurred the development of ensemble decision tree classification methods, such as boosting, bagging, and random forests, as outlined by (Moro et al., 2022). In the domain of "victim detection" analysis, classification algorithms play a crucial role in predicting barriers based on body position patterns (Tham et al., 2021).

Machine learning, as a scientific method, aims to enhance computer learning through experience, as discussed by (Kasnesis et al., 2022). Classification within this framework involves establishing decision rules based on environmental and body orientation characteristics, crucial for tasks like rescue operations. Various classifiers, such as decision trees, neural networks, bat algorithms, artificial bee colonies, particle swarm optimization, support vector machines (SVM), and K-nearest neighbors (K-NN), are commonly utilized in this context.

## IV.VII. Ensemble

Ensemble classifiers, such as random subspace models, combine unrelated subsets of training data or various classifier variables to achieve highly accurate outputs. These classifiers are commonly employed in machine learning, particularly in scenarios like human identification behind barriers in LOS and NLOS fields. By integrating the results from different classifiers, ensemble classifiers make classification decisions effectively (Ayyad et al., 2018).

Ensemble methodologies, also known as ensemble approaches, enhance classification performance by combining outcomes from multiple classifiers. Bagging (bootstrap aggregating) and boosting are prominent strategies within ensemble classification. Bagging involves randomly varying the training data to create replacement training sets while boosting adjusts the weights of training examples based on their influence on classifier performance. Weighted decisions from individual classifiers contribute to creating the final classifier

The method demonstrated by (Kowsari et al., 2019), illustrates the application of boosting algorithms in datasets through ensemble learning, leading to the development of AdaBoost. This iterative approach adjusts weights to improve classifier performance, showcasing advancements in ensemble techniques. Assume that given $Dt$ and ht, it is possible to create $Dt$ such that $D1(i) = \frac{1}{m}$: given $Dt$ and $ht$:

$$D_{i+i}(i) = \frac{D_t(i)}{Z_t} \times \begin{cases} e^{-\alpha t} if\, y_i = h_t(x_i) \\ e^{\alpha t} if\, y_i \neq h_t(x_i) \end{cases} \qquad 1$$

$$= \frac{D_t(i)}{Z_t} exp(-\alpha y_i h_t(x_i)) \qquad 2$$

The normalization variable in this situation is $Z_t$

where
$$\alpha t = \frac{1}{2} in\left(\frac{1 - \epsilon_t}{\epsilon_t}\right) \qquad 3$$

## IV.VIII. Performance Evaluation

Some validation measures are needed to evaluate the effectiveness of the machine learning model. In classification models, the four parameters True Positive (TP), True Negative (TN), False Positive (FP), and False Negative (FN) are typically investigated through confusion matrices (FN). It finds the pictures that were accurately and inaccurately classified from the model dataset provided to assess the model. Figure 2 depicts the Contents of the feature of the loaded raw dataset while performance measurements and how they are calculated are shown below.

A model's accuracy is determined using the four metrics $TP, FP, TN, and\ FN. The\ TP$ product determines the state in which it is present. The result of FP is that it discovers the state when it is absent. Whenever the state is absent, the TN product cannot be located. The state, when it is present, is not found by the FN product.

$$Accuracy: (TP\ +\ TN)\ /\ (TP + TN + FP + FN). \qquad 4$$

Sensitivity is used to determine the proportion of correctly diagnosed cases that have positive positives.

The sensitivity is defined as

$$Sensitivity: (TP\ +\ TN)\ /\ (TP + TN + FP + FN). \qquad 5$$

Specificity determines the number of appropriately recognized occurrences that have negatives.

$$Specificity: TN/\ (FP + TN) \qquad 6$$

$$Precision: TP/\ (TP + FP)\ Recall: TP/TP + FN\ F-Score: 2\ x\ (Recall\ x\ P). \qquad 7$$

## V. Application

Discovery by humans NLOS data processing provides an improved means of discovering victims concealed beneath the debris. The development of several technologies, including automated detection, intrusion detection, and military services, among others, benefits from the requirement to find pertinent data. The use of machine learning technology enables the detection of inconsistent designs and data. It has excellent algorithms, which are tools that are used in many fields. Owing to its simplicity and advantageous programming environment for technologists, designers, researchers, and scholars, among others, MATLAB (Matrix Laboratory) is used for experimenting. MathWorks created MATLAB, a multi-worldview mathematical computational environment, and unique programming language.

It allows for conceptual model controls, function and data charting, algorithm implementation, and the creation of user interfaces in a variety of languages, including C, C++, C#, Java, Fortran, and Python. This article's main focus is on the prediction of victims trapped behind debris utilizing collapse structure technology, which makes use of the MATLAB tool and the NLOS database. An iCore3 CPU, 8 GB of RAM, a 64-bit operating system, and MATLAB 2015b serve as the executing tools on the computer setup used for this investigation.

## VI. Result and Discussion

### VI.I. Result

From NLOS datasets of various data, this study identifies 256 samples per window and 17,408 (static) with 23,552 (dynamic) cases. To standardize the imputed raw data for accurate prediction, SC is mostly utilized for data cleaning. The ICA algorithm was used to extract dormant components from the UWB NLOS data; the ICA feature extraction separates and eliminates discordant parameters to select the determinant discrepancy with a smaller number of independent components to provide meaningful data proof for further investigation. The Ensemble AdaBoost classification algorithm is used to classify the data with recovered ICA latent important features. Using 70% of the data for training and 30% for assessing classification accuracy, ensemble Classifiers are utilized to assess the effectiveness of classification implementation.

The experiment is trained, tested, and evaluated using an ensemble learning approach classification to validate its effectiveness in reducing prejudices in the sample. The computational outcome and performance measures are used to evaluate the results. An AdaBoost ensemble classifier is used to classify the models, and it performs with a performance accuracy of 88.37% and 87.20% for static and dynamic respectively. Figures 3, 4, and 5 below reflect the outcomes and processes. The NLOS data in Table 1 is utilized to extract the hidden features using the ICA feature extraction algorithm. The ensemble technique is used to classify the extracted features, and the outcomes are shown in Table 2, which utilizes the confusion matrix to provide an outcome for the performance metrics in Figure 3.

Figures 3 and 4 successfully categorized and incorrectly classified, are displayed. Numbers and graphs, which also function as measurements for specific data points, signify the values for the variables. This graph shows the correlations between the categorized variables. Figure 3 displays the confusion matrix for the boosted ensemble classifier classifications used in the investigation. The confusion matrix table, which represents the true, false, true negative, and false negative values, is then used to describe the performance of the classification system for groups of tested data with established true values.

| | Max | min | Mean | Std. deviation | Skewness | Kurtosis | Energy | Max / min | Max - min | SD / mean | Max - min sqrd |
|---|---|---|---|---|---|---|---|---|---|---|---|
| 0 | 0.652988 | -0.478481 | -39.062063 | 0.098293 | 1.126915 | 17.370630 | 2.473360 | -1.364710 | 1.131468 | -0.002516 | 1.280221 |
| 1 | 0.680325 | -0.520884 | 0.000047 | 0.104437 | 1.184565 | 16.914978 | 2.792200 | -1.306096 | 1.201208 | 2201.734642 | 1.442902 |
| 2 | 0.654511 | -0.521990 | -78.124979 | 0.104355 | 0.918378 | 15.918480 | 2.787837 | -1.253875 | 1.176501 | -0.001336 | 1.384155 |
| 3 | 0.634057 | -0.541421 | -273.437461 | 0.103709 | 0.991626 | 16.377148 | 2.753444 | -1.171099 | 1.175478 | -0.000379 | 1.381749 |
| 4 | 0.606181 | -0.552172 | 39.062362 | 0.105350 | 0.845794 | 15.462336 | 2.841272 | -1.097813 | 1.158352 | 0.002697 | 1.341780 |

Figure 2. Contents of the feature of the loaded raw dataset

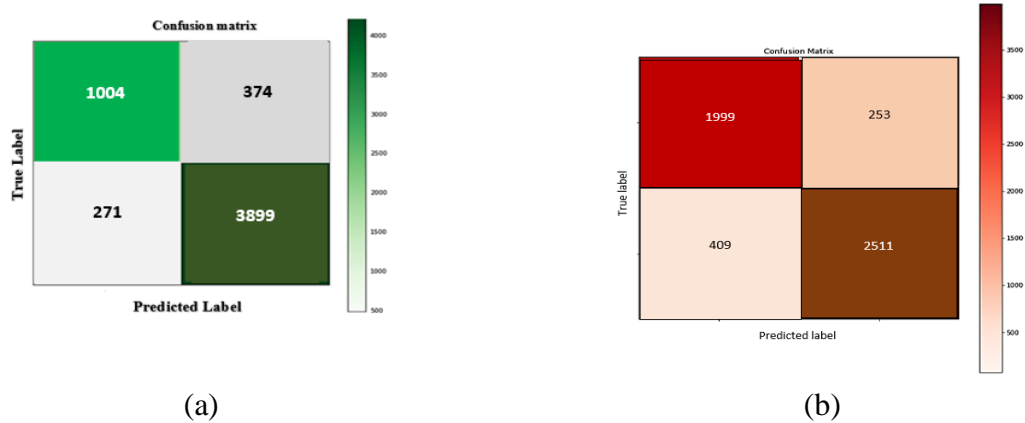

|          (a)          |          (b)          |

Figure 3. Confusion Matrix for the Ensemble Classifications (a) Static and (b) Dynamic loaded data

The data collected was extracted from the measurement data that is accessible for public usage to test the data analysis learning performance methods: https://github.com/disiunibo-nlu/uwb-nlos-human-detection repository.

On 256 observations per window and 17,408 (static) with 23,552 (dynamic) features of the standardized components, the ICA feature extraction method was applied. Performance prediction is carried out using ensemble classification. The outcomes showed how effective data processing is when done using measured data. Table 2 displays and relates the performance outcomes for the suggested strategy. As a result, ensemble categorization performs admirably regarding accuracy and other performance indicators.

The classification of human localization data is investigated more thoroughly in this report. Despite numerous research projects, Figure 4 shows and indicates that a Standard scaler noise reduction model combined with ICA feature extraction approaches can improve ensemble classification outcomes. The performance graph for contrasting output products is displayed in Figure 4. For victims trapped during natural catastrophes, this study suggests a prediction and detection model. The study and performance evaluation of the study findings were displayed in the Figures and Table 2, as previously stated.

Table 2. Performance indicators for the confusion matrix

| Measurement metrics | Static data | Dynamic data |
| --- | --- | --- |
| **Sensitivity** | 78.75 | 83.01 |
| **Specificity** | 91.25 | 90.85 |
| **Precision** | 72.86 | 88.77 |
| **Negative Predictive Value** | 93.50 | 85.99 |
| **Accuracy** | 88.37 | 87.20 |
| **F1 Score** | 75.69 | 85.79 |
| **Matthews Correlation Coefficient** | 67.82 | 74.31 |

The proposed method made use of ICA dimensionality reduction and ensemble classification data analysis operations. The classification of trapped victims of collapsed structural data was analyzed and enhanced in this investigation. Utilizing the performance measures in Figure 4, scholars have proposed several advances, and the results have shown that employing a standard scaler dimensionality reduction approach and ICA feature extraction approaches can improve ensemble classifier accuracy.

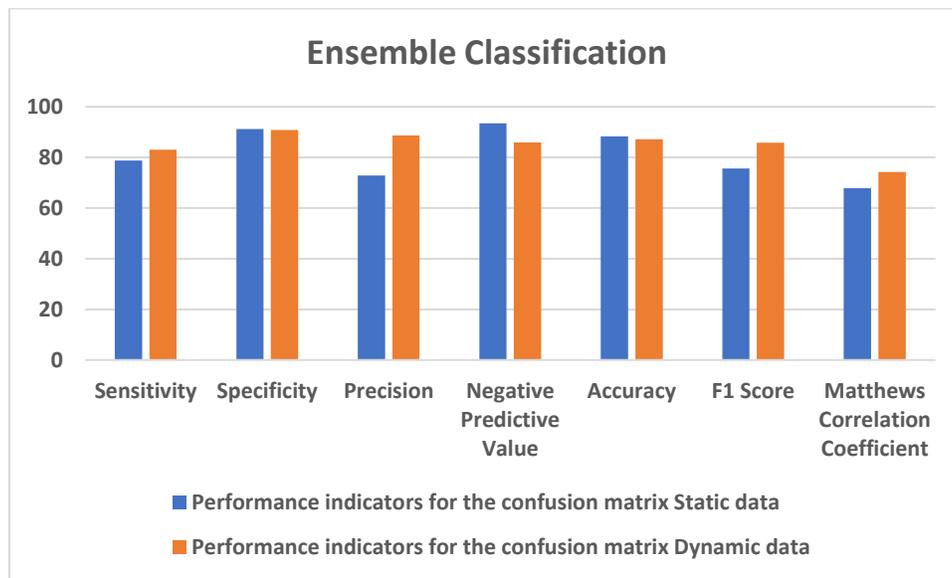

Figure 4: Performance Metrics Graph

## VI.II. Discussion

256 samples per window and 23,552 with 17,408 data of NLOS datasets of various materials were used to construct this study (Moro et al., 2022). The importance of data cleaning, which removes noise and other undesirable qualities that may reduce the output's accuracy, was illustrated using common scales. Reducing the prediction bias introduced by learning algorithms helps machine learning. The subsequent series of categorization experiments that we conducted utilized ensemble learning. The heterogeneous material's pattern sequence only serves to gather the necessary data for categorization using the standard scaler method. SC is applied to the data collected for this investigation, collecting optimal data and reducing uncorrelated features, offering additional detail that is helpful for later research. The attained result is then forwarded to the classifier for algorithm testing and training (Ensemble). The categorization outcomes for the ensemble methodologies are shown. The MATLAB application executes the pattern using ensemble classification techniques.

Accordingly, range-based localization requires the most accurate range predictions, which UWB technology offers (Benouakta et al., 2023). In an indoor context, the NLOS situation complicates the task of location estimation during localization (Ghosh et al., 2020). In LOS situations, UWB can provide an accurate range estimate; in NLOS readings, however, it has issues. The travel time of the received signals lengthens in a non-line-of-sight (NLOS) scenario because of radio wave reflection through scatterers or penetration through barriers that obstruct the signal, like doors and walls.

In the meantime, the NLOS error of each measured TOA needs to be treated as a random variable with a positive bias, which can be rather substantial (Wang et al., 2023). When a robot stays stationary, NLOS detection is rather simple. When the measuring sensor moves, however, it becomes more challenging, which could be one of the reasons why static data evaluation performs better in terms of accuracy and precision (88.37%) than dynamic data evaluation (87.20%).

## VI.III. Validation

Table 3 Comparative approaches

| Methods | Accuracy |
|---|---|
| KNN (Enoch et al., 2023) | 85.00% |
| SVM+Autoencoder (Tran et al., 2022) | 86.98 |
| CNN+stacked-LSTM (Jiang et al., 2020) | 82.14% |

## VII. Conclusion

This research could be useful in identifying people during post-disaster search and rescue operations. This may also result in the use of trustworthy human presence categorization models by software and hardware applications, even in environments with little data. A knowledge-based model was taken from the dataset to categorize whether a person was present or not. The theoretical solution reduces dimensionality using machine learning techniques like modeling and classification algorithms. The ICA filtering function model, which employs ensemble classification, is the approach to dimensionality reduction. This work evaluated and analyzed the study's performance and presented the results of the Ensemble classification method. In this research, the categorization of UWB NLOS human detection signal data was assessed and improved. Study after study has recommended that researchers evaluate their findings via performance metrics. The results have demonstrated that Ensemble classification accuracy can be increased by dimensionality reduction models that use feature extraction techniques like ICA. It is crucial to investigate how the latest suggested study might improve feature extraction models and algorithms. Future research suggests utilizing hybridized dimensionality reduction techniques. Future work on this project can improve the independent component analysis algorithm for better fitness iteration, integrate the strategy with other dimensionality reduction techniques, introduce additional useful classifiers like the KNN and SVM, and then compare and fetch for improved classification and detection efficiency.

## Authors information

[1]Department of Electrical and Information Engineering, Landmark University, Omu-Aran, Kwara State, Nigeria. [2]Department of Electrical and Electronics Engineering, Ekiti State University, Ado Ekiti, Nigeria,

## Competing interests

The authors declare that they have no competing interests.